\documentclass[preprint,12pt,3p]{elsarticle}

\usepackage{amssymb}

\usepackage{subfigure}
\usepackage{indentfirst}     
\usepackage{xcolor,graphicx,float}
\usepackage{eucal}
\usepackage{comment}
\usepackage{multirow}
\usepackage{amsmath}
\usepackage[numbers]{natbib}

\usepackage{geometry}
\journal{Neuralcomputing}

\begin{document}

\begin{frontmatter}

\title{Transfer Learning in General Lensless Imaging through Scattering Media}
    
\author[a,b]{Yukuan Yang\corref{equ}}
\author[c]{Lei Deng\corref{equ}}
\author[a]{Peng Jiao}
\author[d]{Yansong Chua}
\author[a,b]{Jing Pei\corref{c-8082540726b6}}
\author[a,b]{Cheng Ma\corref{c-8082540726b6}}
\author[a,b]{Guoqi Li\corref{c-8082540726b6}}


\cortext[equ]{These authors contributed equally to this work.}
\cortext[c-8082540726b6]{Corresponding author. Email: liguoqi@mail.tsinghua.edu.cn}

\address[a]{ Department of Precision Instrument, Center for Brain Inspired Computing Research\unskip, 
     Tsinghua University\unskip, Beijing\unskip, 100084\unskip, China}
  	
\address[b]{Beijing Innovation Center for Future Chip \unskip, 
    Tsinghua University\unskip, Beijing\unskip, 100084\unskip, China}
  	
\address[c]{Department of Electrical and Computer Engineering\unskip, 
    University of California, Santa Barbara\unskip, CA\unskip, 93106\unskip, USA}
  	
\address[d]{Institute for Infocomm Research (I2R)\unskip, 
    A*STAR\unskip, Singapore\unskip, 138632\unskip, Singapore}


\begin{abstract}
Recently deep neural networks (DNNs) have been successfully introduced to the field of lensless imaging through scattering media. By solving an inverse problem in computational imaging, DNNs can overcome several shortcomings in the conventional lensless imaging through scattering media methods, namely, high cost, poor quality, complex control, and poor anti-interference. However, for training, a large number of training samples on various datasets have to be collected, with a DNN trained on one dataset generally performing poorly for recovering images from another dataset. The underlying reason is that lensless imaging through scattering media is a high dimensional regression problem and it is difficult to obtain an analytical solution. In this work, transfer learning is proposed to address this issue. Our main idea is to train a DNN on a relatively complex dataset using a large number of training samples and fine-tune the last few layers using very few samples from other datasets. Instead of the thousands of samples required to train from scratch, transfer learning alleviates the problem of costly data acquisition. Specifically, considering the difference in sample sizes and similarity among datasets, we propose two DNN architectures, namely LISMU-FCN and LISMU-OCN, and a balance loss function designed for balancing smoothness and sharpness. LISMU-FCN, with much fewer parameters, can achieve imaging across similar datasets while LISMU-OCN can achieve imaging across significantly different datasets. What's more, we establish a set of simulation algorithms which are close to the real experiment, and it is of great significance and practical value in the research on lensless scattering imaging. In summary, this work provides a new solution for lensless imaging through scattering media using transfer learning in DNNs.
\end{abstract}


\end{frontmatter}



\section{Introduction}

\label{sec:Intro}

Experiments show that 83\% of human knowledge acquisition is accomplished via the visual system \cite{Rosenblum2011}. Various imaging tools, such as the camera, telescope and microscope, have been invented to help us see a clearer, more distant, and more microscopic world \cite{Zhuang2016}. However, although human beings have made great progress in imaging technology, there are still many factors that restrict the imaging quality.

The imaging behind or in the scattering media is a kind of pressing problems in many cases, such as the thick tissue imaging \cite{Ma2014,Kang2015}, see around the corner \cite{Gariepy2015}, fog imaging \cite{Psaltis2012} and light focus in the complex media \cite{Horstmeyer2015}. Due to the actions of molecules or atoms in the medium, light intensity, propagation direction, polarization state or even frequency will change in the propagation of light. Up to now, there is still no effective and simple solution to solve the inverse problem for imaging behind the scattering media. The strong scattering will occur when the light waves propagate in random scattering media and generate a random pattern, called speckle. It always requires a huge amount of measurements and sophisticated controlling to characterize approximately the scattering media.

At present, the common imaging through scattering media methods include wavefront modulation, transmission matrix, and speckle correlation. The wavefront modulation method \cite{Yaqoob2008,Mosk2012,Vellekoop2010,VanPutten2011} is used to compensate for the change in amplitude or phase of the incident light waves by using the pixel-modulated spatial light modulator (SLM) or the digital micromirror device (DMD) to reduce the impact of the scattering and thus improve imaging quality. However, the wavefront modulation method needs precise feedback control \cite{Shao2013,He2013,Wu2014}, and the imaging quality is hard to be improved with the limitation of the SLM or DMD partition density. The transmission matrix method can reconstruct the image by using the transmission matrix (TM) \cite{Kim2015,Vellekoop2007}, which can describe the characteristics of the random scattering media accurately. The fixed scattering medium is regarded as a linear shift-variant matrix. Due to the high susceptibility of model errors resulting from speckle decorrelation \cite{Hillman2013,Liu2015,Qureshi2017}, a slight perturbation of the scattering process could lead to much-reduced correlations between the speckles measured before and after. And TM is inevitably large, resulting from the many underlying degrees of freedom. Both reasons lead to complex computing and poor imaging quality. Speckle correlation method \cite{Freund1990} utilizes the correlation among speckle regions within the range of optical memory effect \cite{Freund1988}, which approximates the system to be a shift-invariant system. But the method is limited by the small FOV \cite{Schott2015} and finite sensor dynamic range \cite{Katz2014}, rigorous illumination coherence and measurement requirement \cite{Bertolotti2015}.

In recent years, deep neural networks (DNNs) \cite{Lecun2015} have made a huge impact in the field of artificial intelligence. It has achieved the state-of-the-art results in face recognition \cite{Parkhi2015}, image restoration (e.g, colorization \cite{Cheng2015}, deblurring \cite{Dong2014,Xu2014,Sun2015}, and inpainting \cite{Xie2012}), speech understanding \cite{Gavat2015}, text analysis \cite{Santos2014}, robotics \cite{zhang2016}, and so forth. With a deeper understanding and increased processing speed, DNNs have also begun to venture into other domains. The feature extraction capability of DNNs may help other fields with problems that cannot be analytically addressed using conventional methods. For example,
DNNs have been applied to the field of imaging through scattering media \cite{Ando2015,Meng2017,Sinha2017}. Ando et al. demonstrated the object recognition of face and non-face images through the corresponding speckle intensity images by using support vector machines (SVM) \cite{Ando2015}. Lyu et al. used the multi-layer perceptron to realize imaging through scattering media on handwritten digits and letters \cite{Meng2017}. Sinha et al. introduced the convolutional neural network (CNN) to lensless imaging through scattering media by training it on the more complex dataset, namely ImageNet \cite{Sinha2017}. Although the above work has achieved good results, a common problem persists. The model trained on one dataset cannot be used for a different dataset. Hence, if we want to perform imaging on a new dataset, we would need another tens of thousands of speckle intensity images from that dataset for training from scratch. Both the data collection and training of a new DNN are very time consuming and also costly.

Noting that lensless imaging through scattering media is a high dimensional inverse regression problem, the DNN trained on one dataset, in general, cannot be used to recover the images from other datasets. Fortunately, the scattering process is physically similar when the scattering media and parameters are the same or similar. Therefore, a DNN trained using a relatively more complex and hence more general dataset may be reused to reduce the training time and samples required for training on a similar or simpler dataset. Transfer learning \cite{Torrey2010}, one of the most important research fields in deep learning \cite{Lecun2015}, whose goal is to enhance the generalizability of a trained DNN across different datasets, could be applied to generalize a trained DNN across different datasets. For example, based on the model trained using the CIFAR-10 \cite{CIFAR10_dataset} dataset, we just need several hundreds of samples from other datasets such as the MNIST \cite{lecun1998gradient}, Fashion-MNIST \cite{xiao2017fashion} and Face \cite{1994Parameterisation} dataset, to re-train the model. In contrast, we may need tens of thousands of training samples if we are to start from scratch. By doing so, the number of samples from the new dataset and the time required for retraining are reduced by two orders of magnitude. {Due to the huge cost of data acquisition, we establish a set of simulation algorithms to simulate the scattering process.} Specifically, considering the difference in sample sizes and similarity among datasets, we propose two DNN architectures, namely, LISMU-FCN (Lensless Imaging through Scattering Media U-Fully Convolutional Neural Network) and LISMU-OCN (Lensless Imaging through Scattering Media U-Ordinary Convolutional Neural Network), which can be used in different scenarios. When the datasets are similar, LISMU-FCN can realize real-time imaging with fewer parameters for training and inference; while LISMU-OCN has better transfer learning capability and can be used across significantly different datasets. Besides, to keep the sharpness of images output from LISMU-FCN and remove the noise of images output from LISMU-OCN, a novel balance loss function based on the Sobel and Laplacian operator is proposed to balance the smoothness and sharpness of the output images. Our work hence introduces transfer learning in DNN to lensless imaging through scattering media. {To sum up, our contributions mainly include the following three points: 

(a) Aiming at the problem of complicated experimental conditions and requirements in data construction, we have established a set of simulation algorithms that can closely mimic the physical generation, which is of great significance and practical value in the research on lensless scattering imaging.

(b) We are the first that introduces transfer learning into the field of lensless imaging through scattering media to solve the problem of costly data acquisition, reducing the amount of data needed for model training by two orders of magnitude. Moreover, we have analyzed the training dataset selection in transfer learning and its importance for imaging quality.

(c) We propose a novel balance loss function to improve the performance by keeping the sharpness and reducing the noise of the images.

}

\section{Lensless Imaging through Scattering Media System}

\begin{figure}[htbp]
    \centering
    \includegraphics[height=0.22\textwidth]{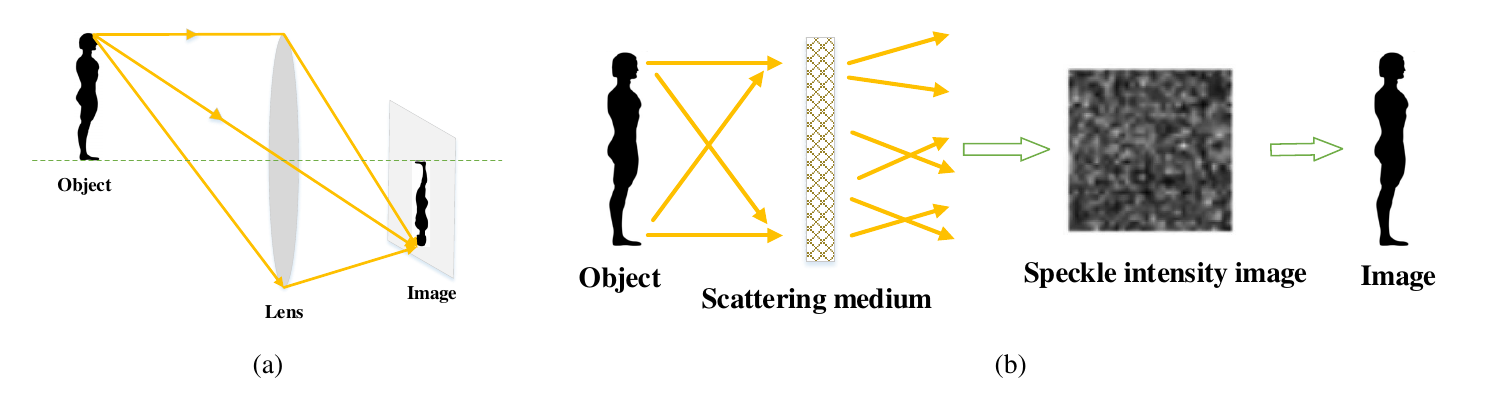}
    \caption{(a) Illustration of lens imaging; (b) Illustration of lensless imaging through scattering media.}
    \label{fig:1}
\end{figure}

The schemes of the traditional lens imaging system and imaging through scattering media system are as shown in Figure \ref{fig:1}. Different from the traditional lens imaging system, imaging through scattering media leverages on the diffraction of light waves so as to realize lensless imaging. Traditional lens imaging usually uses the lens to image the target object onto the photographic plate. However, in lensless imaging through scattering media system, the scattering media can be regarded as an unknown random mask embedded in the imaging system which scrambles the wave front and projects a speckle intensity image onto the imaging plane. In order to recover the original image successfully, it is important to characterize the random mask as well as possible. The transmission matrix (TM) method characterizes it by modulating the different modes of the input light \cite{Kim2015}. However, a sophisticated and expensive modulation system is then required. The precision of characterization is also affected by noise in the acquisition process. Even if the characterization of the mask is complete, the reconstruction, which is the inverse problem, is an ill-conditioned problem. In this paper, we would recover the target image behind the scattering media using DNNs techniques.

\begin{figure}[htbp]
    \centering
    \includegraphics[width=0.48\textwidth]{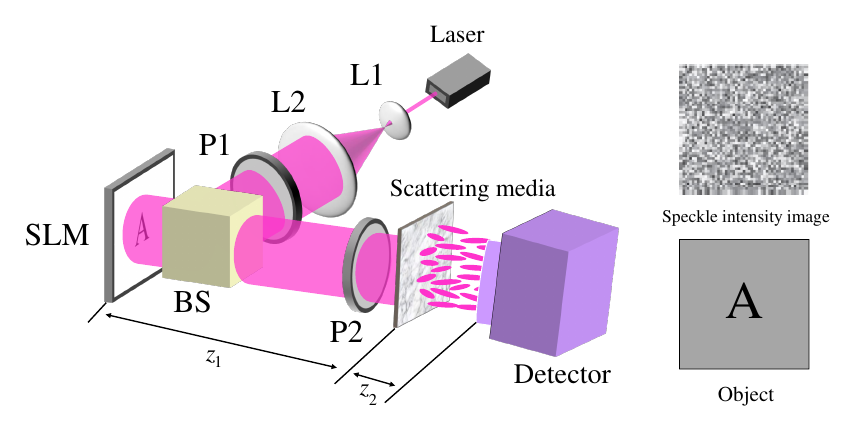}
    \caption{The optical setup for lensless imaging through scattering media system.}
    \label{fig:2}
\end{figure}

The optical setup of imaging through scattering media system is illustrated in Figure \ref{fig:2}. The laser beam is filtered by spatial filter pinhole (L1) and then collimated by the lens (L2). The plane wave passes through a linear polarizer (P1) and is reflected by the beam splitter (BS). The incident beam is modulated by an amplitude-only spatial light modulator (SLM) that serves as the target image.

The reflected and modulated beams go through the beam splitter and the linear polarization analyzer (P2) that enhances the intensity modulation contrast. The light from the target image propagates at the free space and scrambled by the scattering media which is regarded as ground glass. The distance from the target image to the scattering media is $z_{1}$. A detector is placed behind the scattering media to record the speckle intensity. The distance from the scattering media to the detector is $z_{2}$.

The ground glass could be approximated as a complex-amplitude modulator whose complex transmittance $t(x,y)$ at position $(x,y)$ is described as

{
\begin{equation}
\centering
t(x,y)=A(x,y)exp{ \left[ \frac{2\pi\Delta n}{\lambda}i \cdot D(x,y) \right]}
\label{equ:1}
\end{equation}}

\noindent where $A(x,y)$ is the amplitude modulation of the ground glass, and $\Delta n$ is the difference between the refractive index of the ground glass and the surrounding medium. The ground glass is placed in the air, hence $\Delta n$ is set to 0.52 in the simulation. $\lambda$ is the wavelength of the incident monochrome light which is set to 633nm. $D(x,y)$ is the random surface height of the glass diffuser, which is expressed as the phase modulation. The output complex field $U_{out}$ at the image plane is expressed as a function of the input complex field $U_{in}$ at the object plane using the angular spectrum method (ASM) operator $\psi$ in

{
\begin{equation}
\begin{split}
U_{out}&=\psi(U_{in})=F^{-1} ( \hat U_{in}exp{ \left[ 2\pi z i \cdot \sqrt{{(\frac{1}{\lambda})}^{2}-f_{x}^{2}-f_{y}^{2}]}  \right] })
\end{split}
\label{equ:2}
\end{equation}
}

\noindent where $\hat U_{in}$ is the angular spectrum of $U_{in}$, which is also expressed as the Fourier transform of $U_{in}$, while $F^{-1}$ is the inverse Fourier transform, and $f_x$, $f_y$ are spatial frequencies in the direction of $x$ and $y$ in the object plane. $z$ is the distance from the object plane to the image plane. The speckle intensity $I(x,y)$ at position $(x,y)$ recorded by the detector is then given by
{
\begin{equation}
I(x,y)=H (g(x,y)) = \left| \psi_{z_{2}}  \left\lbrace \psi_{z_{1}} \left[ g(x,y) \cdot t(x,y) \right]  \right\rbrace \right| ^2
\label{equ:3}
\end{equation}
\noindent  where $t(x,y)$ is the complex transmittance of the ground glass defined in Equation (\ref{equ:1})}, and $g(x,y)$ is the target image displayed at the SLM, and $H$ is the forward operator that maps $g(x,y)$ to $I(x,y)$. In order to reconstruct the distribution of the target image, the inverse operator $H^{-1}$ has to be obtained. However, it is difficult to characterize the random scattering media and hence the explicit form of $H^{-1}$. Even if the random scattering media could be characterized completely, computing $H^{-1}$ is costly, given the large size of $H$.

Here, we use DNNs to approximate the inverse effect of the scattering media by training a DNN with a large amount of data, using speckle intensity and target images. With the help of deep learning, lensless imaging through scattering media is treated as a high dimensional regression problem. Convolutional neural network (CNN) with stochastic gradient descent (SGD) is a good choice for this problem. In addition, since CNN is data-driven, it does not require any additional information about the scattering process other than the images. However, this also necessarily means that a large amount of data, sometimes even tens of thousands of samples, is required. Transfer learning is hence employed to solve this problem, which is described in Section \ref{sec:transfer_learning}.

As DNNs require a lot of data for training and inference, the data is generated by simulation according to Equation (\ref{equ:3}) on several common image datasets which are easily accessible large-scale image datasets used in deep learning, e.g. MNIST, CIFAR-10, Fashion-MNIST, Face dataset, etc. {Accessibility and complexity are the criterions for the training dataset selection} and the original target images contain $32 \times 32$ pixels. They are resized into $140 \times 140$ pixel images and then zero-padded on all four sides to create $800 \times 800$ pixel images to satisfy the sampling condition of ASM. {Although the scaling and zero-padding process increases the amount of data computation, it can bring the following advantages, such as increasing the light propagation distance, making the simulated speckle intensity images closer to the real speckle intensity images, extending the field of view because of the enlarged uploading picture, etc.} The processed images are displayed in the SLM with size $8\mu m \times 8 \mu m$. The scattering media is considered as a complex-amplitude modulator and hence represented using an $800 \times 800$ random complex matrix.

The amplitude and phase of the matrix are sampled from uniformly distributed random variables ranging from 0 to 1 and 0 to $2\pi$, respectively.
{The reason why we choose the uniform random distribution for amplitude and phase of matrix lies in that the uniform random distribution provides an extreme scrambler for the light modulator, which could provide the maximum degree of freedom to modulate the light. That is to say, the uniform random distribution provides the biggest difficulty for imaging and can be used to verify that the proposed method based on deep learning has a powerful fitting ability to reconstruct the object through the scattering media.}
The distance $z_1$ and $z_2$ are set to $20mm$ and $10mm$, respectively. And the speckle intensity images recorded by the detector are of pixel size $800 \times 800$, which we then crop at the center to $32 \times 32$ for training.

\begin{figure}[htbp]
    \centering
    \includegraphics[width=0.96\textwidth,height=0.25\textwidth]{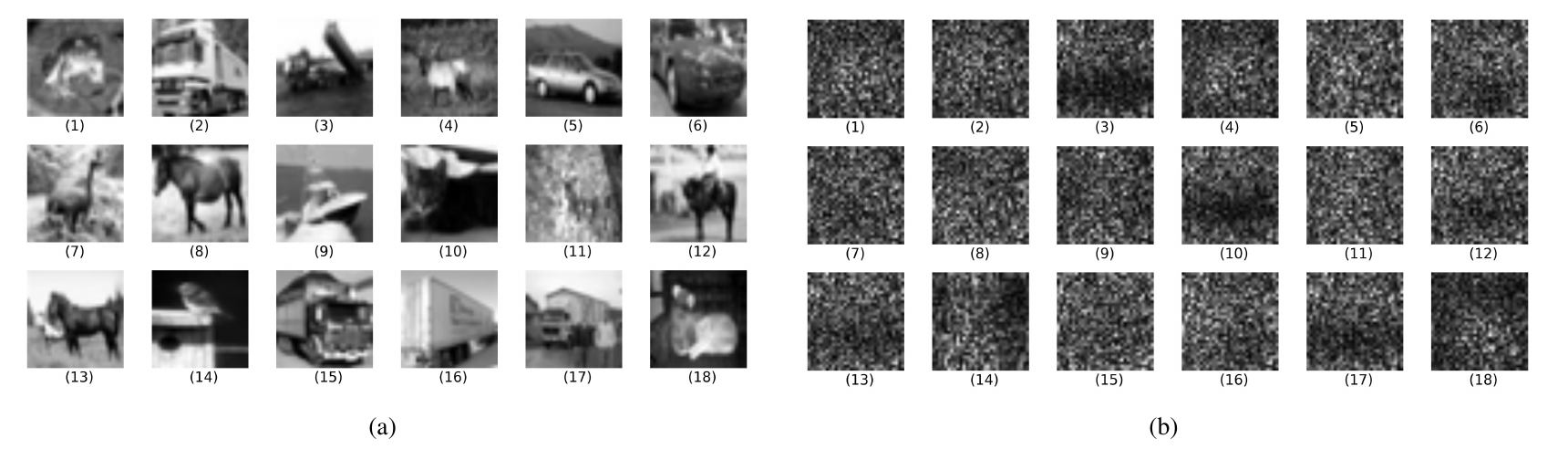}
    \caption{Samples from CIFAR-10 \cite{CIFAR10_dataset} : (a) imaging targets; (b) corresponding speckle intensity images.}
    \label{fig:3}
\end{figure}

Here we use CIFAR-10 to illustrate the scattering results of the simulation. CIFAR-10 includes 10 object classes such as aircraft, car, ship, bird, cat, dog, etc. 50,000 images are used for training with another 10,000 used for testing. We use the above simulation algorithm to image all 60,000 images from CIFAR-10 so as to obtain the corresponding speckle intensity images. Figure \ref{fig:3} illustrates the imaging targets (ground truth) from CIFAR-10 and the resulting speckle intensity images.

\section{Convolutional Neural Networks}

Inspired by the biological visual system, the first CNN was proposed by LeCun in 1989 for recognizing handwritten digits\cite{Lecun2014}. It went on to outperform all other methods in the large-scale image classification task based on the ImageNet dataset with significant accuracy improvement in 2012 \cite{Krizhevsky2012}. Since then, CNN has been widely used in various fields. With local receptive fields for efficiently extracting local features and shared weights for greatly reducing the number of training parameters, hence avoiding overfitting, CNN performs much better in solving classification and regression problems.

\subsection{Network Architecture}

Based on the acquired speckle intensity images, we design the CNN model for recovering the original images. Inspired by the classic network architecture of U-net in the field of image recovery, we design a CNN that matches the data characteristics of our use case. The U-net \cite{Ronneberger2015} architecture is initially used to detect the edges of cell pictures. 
It consists of a contracting path to capture context and a symmetrical expanding path to enable precise localization. Such a network can be trained end-to-end using very few images and won the ISBI cell tracking challenge 2015. As the initial downsampling and then upsampling architecture is analogous to the process of extracting image features and then recovering them, it is naturally favored by researchers in the field of image recovery. The cross-layer connections result in better convergence, making training more efficient.

\begin{figure*}[htbp]
    \centering
    \includegraphics[width=0.95\textwidth]{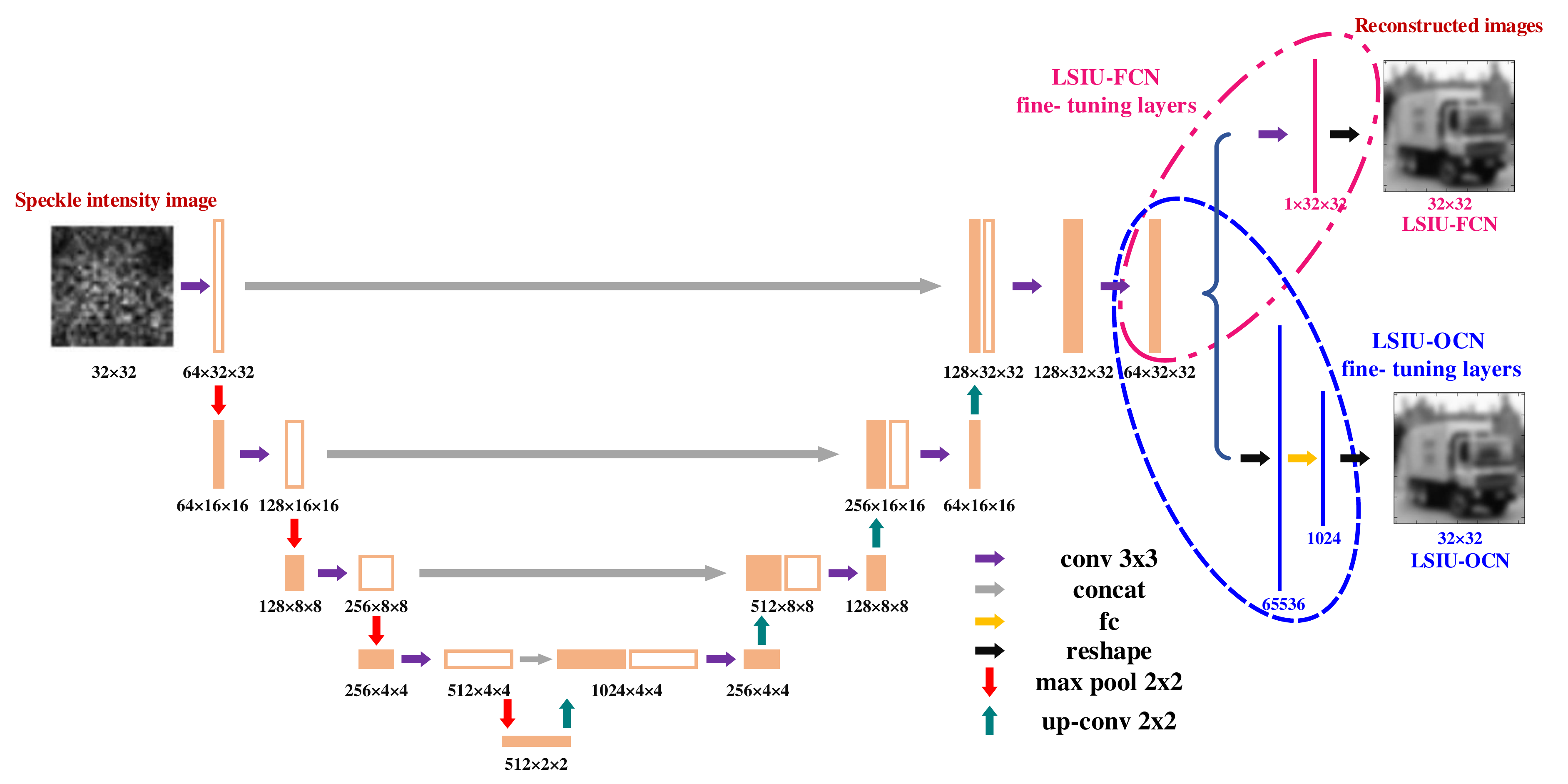}
    \caption{Network structures of LISMU-FCN and LISMU-OCN. The speckle intensity image of the image from CIFAR-10\cite{CIFAR10_dataset} is obtained by the simulation model, which is used to simulate the real scattered image. The reconstructed images are recovered from the speckle intensity image through the DNN models.}
    \label{fig:4}
    \vspace{-15pt}
\end{figure*}

Similar to the U-net, we design two network architectures for lensless imaging through scattering media, namely the LISMU-FCN and LISMU-OCN.
{The U-net model is originally used for semantic segmentation and here we adapt it for image construction and confirm the effectiveness in the field of imaging through scattering media. And we have explored the best depth and parameter quantity of the network according to the dataset since a deeper network will waste computing resources and need more training data while a shallower network will cause the poor imaging quality. Moreover, we also add a full connection layer in LISMU-OCN to increase the capacity of transfer learning.}
The architectures of LISMU-FCN and LISMU-OCN are presented in Figure \ref{fig:4}. The only difference between LISMU-FCN and LISMU-OCN is that the last layer of LISMU-FCN is still a convolutional layer while LISMU-OCN has a fully connected layer instead. Due to the parameter-intensive fully connected layer, LISMU-OCN has thousands of times more parameters than LISMU-FCN, which endows LISMU-OCN with stronger transfer learning capability. However, this also means that it requires more training samples and training time for convergence.

\subsection{Balance Loss Function}
As aforementioned, lensless imaging through scattering media is a high dimensional regression problem. To solve this, we define a new loss function for optimizing the CNN. In order to make the output images as close as possible to the ground truth while balancing the smoothness and sharpness of the images, we add a balance loss term to the mean square error (MSE). To avoid overfitting, an L2-norm regularization term is also used. The whole loss function is calculated by

{
\begin{eqnarray}
\begin{split}
&~~~loss=mse_{loss}+balance_{loss}+L2_{loss}\\
&mse_{loss}=\sum_{i}^{M}\sum_{j}^{N}[ y_{output}(i,j)-y_{real}(i,j) ]^2 \\
balance_{loss}&=\lambda [ \| h*y_{output} \|_2^2 + \| h^T*y_{output} \|_2^2 + \| l*y_{output}\|_2^2 ]\\
&~~~~h=\left[
\begin{matrix}
-1 & 0 & 1 \\
-2 & 0 & 2 \\
-1 & 0 & 1
\end{matrix}    
\right],~~~
l=\left[
\begin{matrix}
0 & 1 & 0 \\
1 & -4 & 1 \\
0 & 1 & 0  \\
\end{matrix}
\right] \\
&~~~~~~~~~~~~~~~~~~~~~~L2_{loss}= \sigma \sum_{w} w^2
\end{split}    
\label{equ:4}
\end{eqnarray}
}

\noindent where $y_{real}(i,j)$ is the gray value of the pixel at $i$-th row and $j$-th column in the ground truth and $y_{output}(i,j)$ denotes the corresponding gray value from output image, and $M$, $N$ are the image width and height. Specifically, in the balance loss term, $h$ and $l$ are the Sobel and Laplacian operators for extracting the image gradient and $\lambda$ is a balance constant. Positive $\lambda$ is able to retain the image smoothness while a negative $\lambda$ retains the image sharpness. In addition, $L2_{loss}$ is the L2-norm regularization on the model parameters for preventing overfitting, and $w$, $\sigma$ are the weight in DNNs and a small constant, respectively.
 

\section{Transfer Learning}
\label{sec:transfer_learning}

    Although DNNs have made much progress, their tedious training process \cite{Yosinski2014} and poor generalization \cite{Neyshabur2017} are key areas of ongoing research. Most DNN models work well only under the assumption that the training and inference data are drawn from the same feature space and have the same distribution \cite{Pan2010}. If the distribution changes or the samples are from different datasets, the model needs to be retrained from scratch over newly collected training data.

Lensless imaging through scattering media is a high dimensional regression problem whereby in general it is difficult to obtain an analytical solution. A DNN trained on one dataset cannot be used in general to recover images from other datasets. To avoid collecting huge data for each new dataset and retraining from scratch, transfer learning is applied to reduce the training samples and training time required for re-training on a different dataset. 
The prior knowledge acquired from the source task is used for the target task. It works by first training a DNN in a relatively complex dataset using a large number of training samples and then fine-tuning the last few layers based on the few samples from each specific new dataset. For example, based on a DNN trained on CIFAR-10, we require very few samples, just a few hundreds, from other datasets such as MNIST, Fashion-MNIST, and Face dataset, to fine-tune the model to generate target images from these new datasets, while training from scratch typically requires tens of thousands of training samples.

\subsection{Transfer Learning Definition and Categorization}
\begin{figure*}[htbp]
    \centering
    \includegraphics[width=1.0\textwidth]{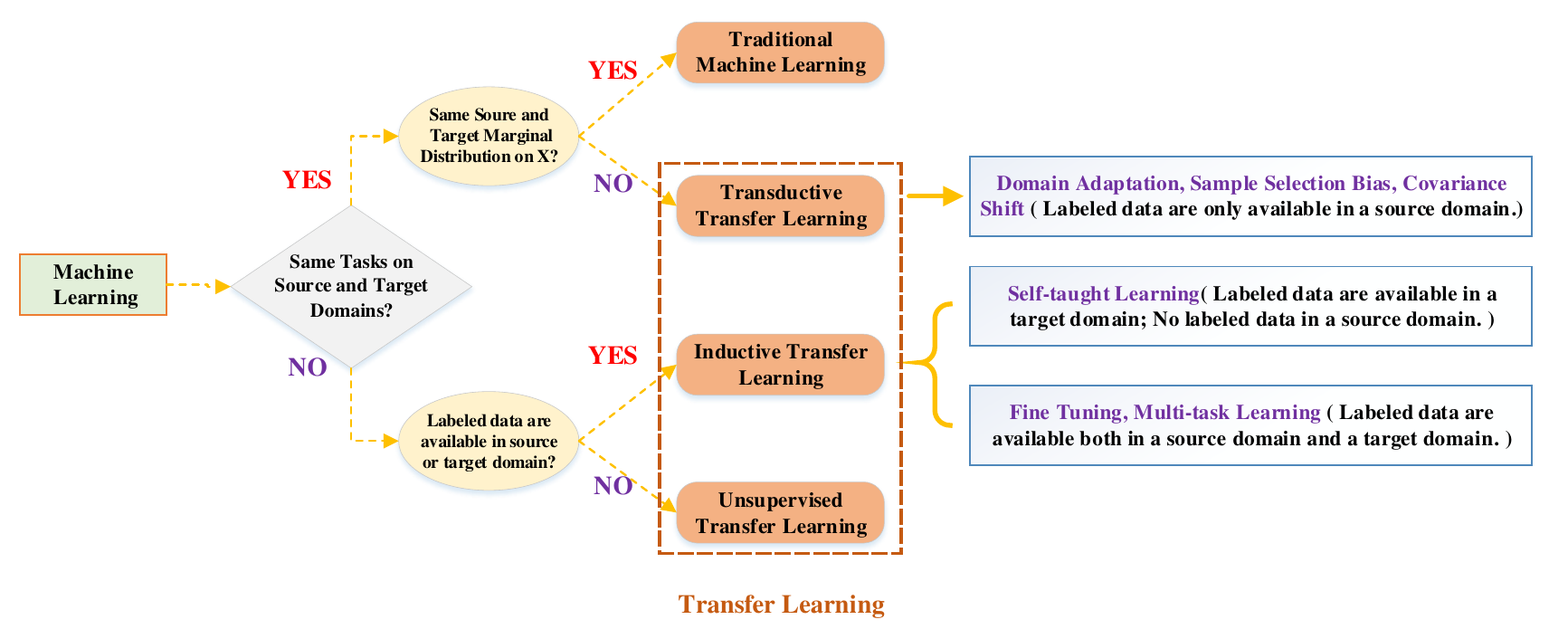}
    \caption{Transfer learning categorization.}
    \label{fig:5}
    \vspace{-15pt}
\end{figure*}

\vspace{10pt}

Transfer learning\cite{Torrey2010} is one of the most important research fields in deep learning, of which its goal is to enhance the generalizability of a trained DNN across different datasets. To formally define transfer learning, we first introduce the below notations and definitions\cite{Pan2010}.
Domain $\mathcal{D}=\{\mathcal{X},P(X)\}$ is defined as a set consisting of feature space $\mathcal{X}$ and its corresponding marginal probability distribution $P(X)$, where $X=\{x_1,x_2,...,x_n\}\in \mathcal{X}$. Task $\mathcal{T}=\{\mathcal{Y}, f(\cdot)\}$ is defined as a set consisting of label space $\mathcal{Y}$ and a predictive function $f(\cdot)$, in which $f(\cdot)$ is learned from training data that contains pairs $(x_i,y_i)$, where $x_i\in \mathcal{X}, y_i\in \mathcal{Y} $. Given the definitions of domain $\mathcal{D}$ and task $\mathcal{T}$, transfer learning is defined as follows. Given a source domain $\mathcal{D_S}$ and learning task $\mathcal{T_S}$, and a target domain $\mathcal{D_T}$ and learning task $\mathcal{T_T}$, transfer learning aims to improve the performance of the target predictive function $f_T(\cdot)$ in $\mathcal{D}_T$ using a DNN trained based on $\mathcal{D}_S$ and $\mathcal{T}_S$, where $\mathcal{D}_S \neq \mathcal{D}_T$ or $\mathcal{T}_S \neq \mathcal{T}_T$.

Based on the source and target domains and tasks, transfer learning can be categorized into three sub-sets: inductive transfer learning, transductive transfer learning, and unsupervised transfer learning \cite{Pan2010}. Figure \ref{fig:5} illustrates the different categories of transfer learning. This work belongs to the domain of inductive transfer learning. Here the labeled data of source and target domains are available. Although the amount of labeled data from target domains is too small to train a new model from scratch, it is enough for model fine-tuning given the knowledge transferred from the source domain and task.

\subsection{Model Fine-tuning}

There are many factors that determine the transfer learning strategy to be used, of which two of the most important ones are the size of the new data available and its similarity to the dataset used for model training. Considering the high cost of data acquisition, we would minimize the number of samples required from the new dataset for model retraining. It is also critical that a relatively complex dataset is used for model training to guarantee the similarity between the re-training and training datasets. {In the task of lensless imaging through scattering media, the training dataset selection is a key factor because whether the training DNN model is the same or close to the real imaging model indeed matters. The dataset selection determines the task difficulty and effectiveness of transfer learning. Because the speckle images of the complex dataset with more texture details and gradient information can better reflect the scattering process, the data can help the DNN model to get closer to the real model and have a better transfer learning ability. And this is the reason why we use CIFAR-10 rather than MNIST to train the DNN model.}

Fine-tuning is an important transfer learning strategy for situations when the target data is limited and the learning tasks are related. Observation of CNNs shows that the front layers contain more generic features while the last layers are more dataset-specific \cite{Yosinski2014}. 
Using a pre-trained CNN on source dataset with a large sample size and high complexity, the parameters in the front layers are retained and only the last layers are retrained using the target dataset. The fine-tuning of LISMU-FCN and LISMU-OCN are depicted in Figure \ref{fig:4}. We simply fine-tune the last two layers of LISMU-FCN and LISMU-OCN using a small number of samples from target datasets.

\section{Experiment Results}
\subsection{Model Training and Inference on CIFAR-10 Dataset}

\begin{figure}[htbp]
    \begin{minipage}[t]{0.45\textwidth}
        \centering
        \includegraphics[height=0.48\textwidth]{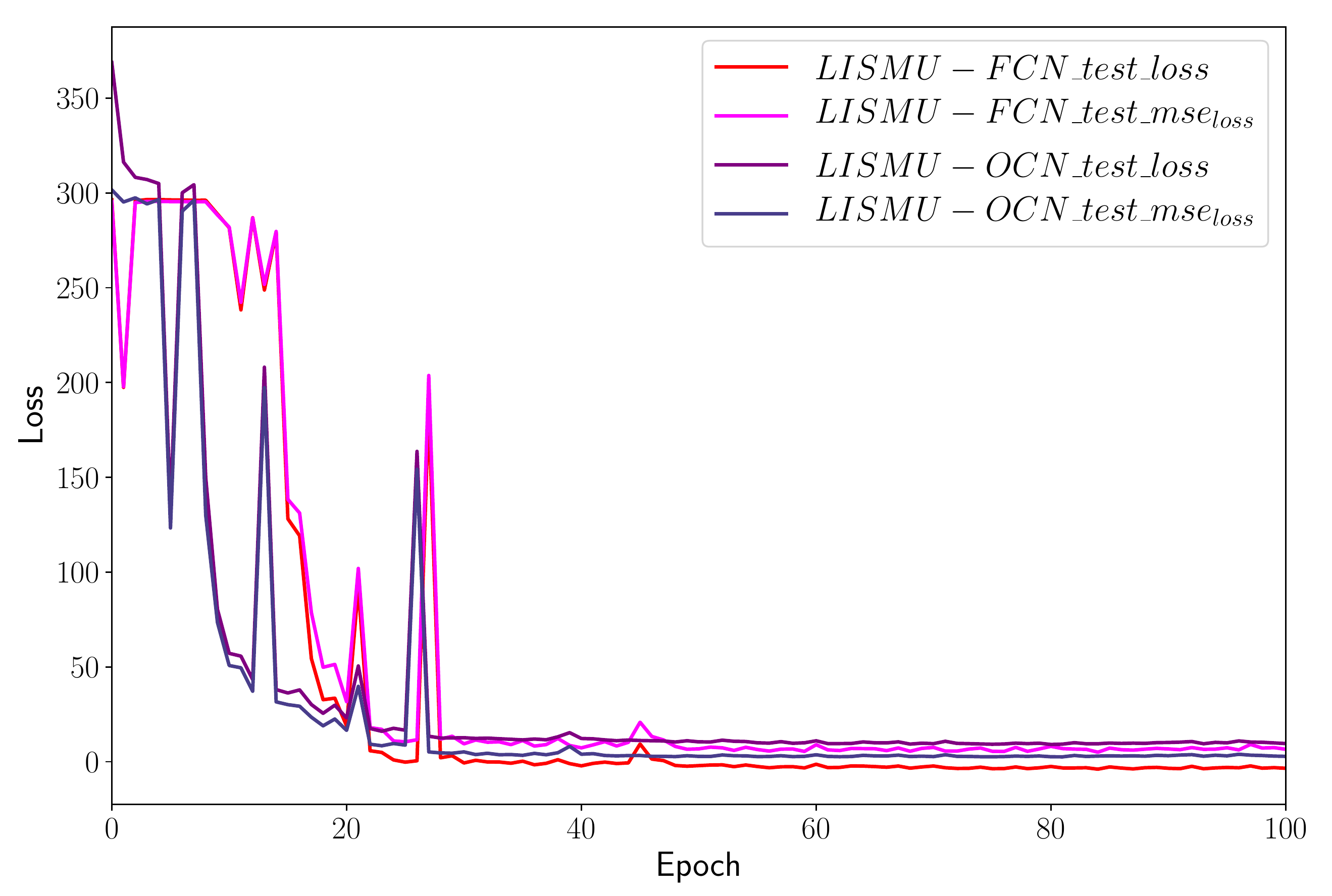}
        \caption{{LISMU-FCN and LISMU-OCN loss curves.}}
        \label{fig:6}        
    \end{minipage}
    \hspace{10pt}
    \begin{minipage}[t]{0.45\textwidth}
        \centering
        \includegraphics[height=0.50\textwidth]{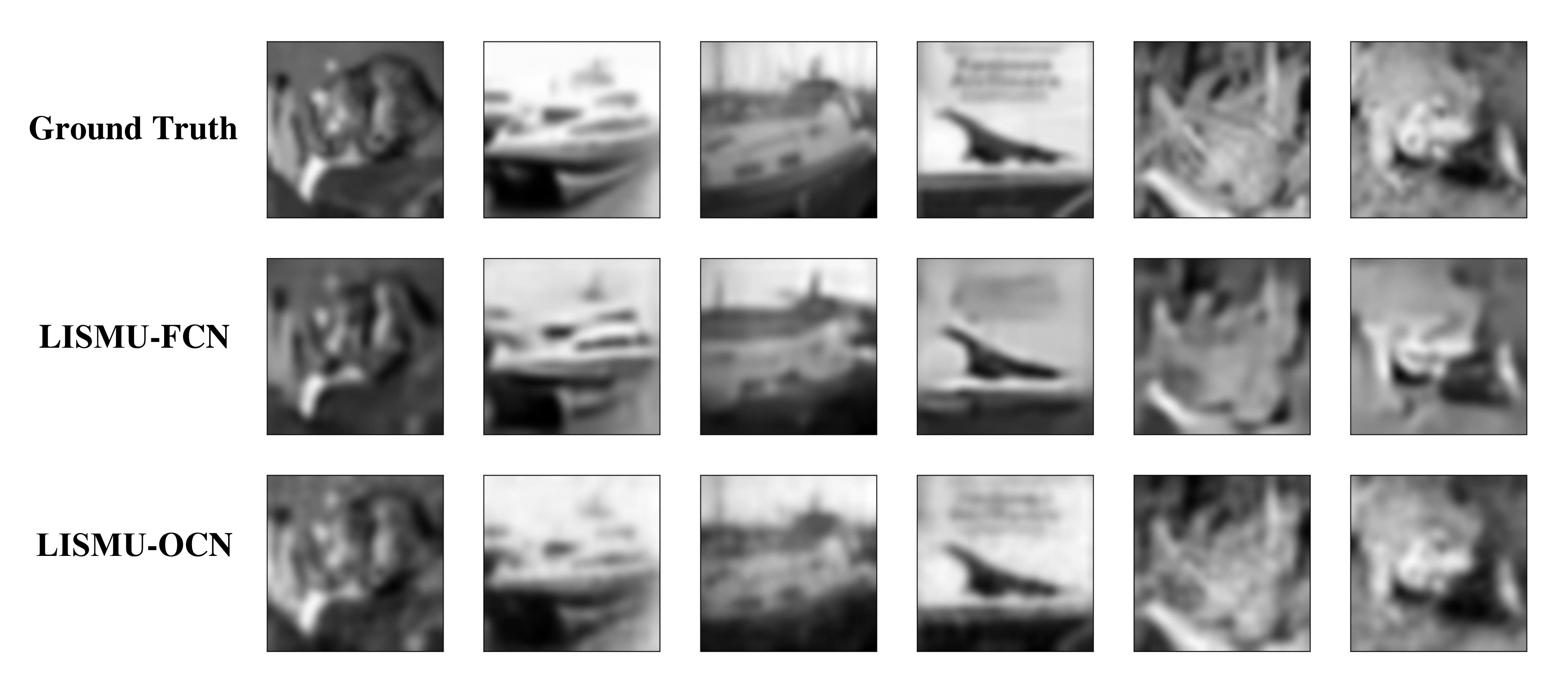}
        \caption{Ground truth (CIFAR-10\cite{CIFAR10_dataset}) and output images from LISMU-FCN and LISMU-OCN.}
        \label{fig:7}
    \end{minipage}    
    
\end{figure}

Since CIFAR-10 is a relatively complex dataset (detailed analysis can be found in Subsection \ref{sec:exp-transfer}), it is used to train our model. Figure \ref{fig:6} illustrates the loss values as a function of the number of steps during the training of LISMU-FCN and LISMU-OCN. Among all loss terms, $mse_{loss}$ is the one to measure the similarity between model output images and ground truth. 

The imaging results on the CIFAR-10 test set are shown in Figure \ref{fig:7}. It is difficult to distinguish between the output images of LISMU-FCN and LISMU-OCN using the naked eye. In order to evaluate the imaging quality more precisely, we introduce four measures, namely, the average gray difference, the peak signal to noise ratio (PSNR), the structure similarity (SSIM) \cite{SSIM} and the classification accuracy.

The pixel-wise average gray difference between ground truth and output images is described as
\begin{equation}
|\bigtriangleup x |= \dfrac{1}{MN} \sum_{i}^{M} \sum_{j}^{N} | y_{real}(i,j)-y_{output}(i,j) |
\label{equ:5}
\end{equation}
\noindent where $y_{real}(i,j)$ is the gray value of the pixel at $i$-th row and $j$-th column in the ground truth, $y_{output}(i,j)$ denotes the corresponding gray value from output image, and $M$, $N$ are the image width and height.

Peak signal to noise ratio (PSNR) is an objective criterion for image evaluation. It is widely used in the image processing field to measure similarity between images. Here we use it as an evaluation of imaging recovery quality. PSNR is formulated as
\begin{eqnarray}
\begin{split}
MSE&= \sum_{i}^{M}\sum_{j}^{N}|y_{real}(i,j)-y_{output}(i,j)|^2\\
&~~~~~PSNR=20log_{10} \frac{MAX_I}{\sqrt{MSE}}
\end{split}
\label{equ:6}
\end{eqnarray}
\noindent where $MAX_I$ is the maximum gray value of the images, which is $255$ here.

{In addition, to evaluate the imaging quality more comprehensively, we also introduce structure similarity (SSIM) as the evaluation index, which measures image similarity from three aspects: luminance, contrast and structure. SSIM is defined as

\begin{equation}
SSIM=\dfrac{(2\mu_{y_{real}} \mu_{y_{output}}+c_{1})(2\sigma_{y_{real}y_{output}}+c_2)}{(\mu_{y_{real}}^2+\mu_{y_{output}}^2+c_1)(\sigma_{y_{real}}^2+\sigma_{y_{output}}^2+c_2)}
\end{equation}

\noindent where $y_{real}$ and $y_{output}$ are the ground truth and the output images of DNNs, respectively. ($\mu_{y_{real}}$, $\mu_{y_{output}}$) are the mean value of ($y_{real}$, $y_{output}$), and ($\sigma_{y_{real}}$, $\sigma_{y_{output}}$) are the standard deviation of ($y_{real}$, $y_{output}$). $\sigma_{y_{real}y_{output}}$ is the covariance of $y_{real}$ and $y_{output}$, and ($c_{1}$, $c_2$) are constants.

\begin{table}[!htbp]    
    \begin{center}
        \begin{tabular}{|c|c|c|c|}
            \hline
            Network & $|\bigtriangleup x|$ & $PSNR~(dB)$  & $SSIM$\\
            \hline
            LISMU-FCN & 12.75 & 24.00  & 0.81\\
            \hline
            LISMU-OCN & 9.34 & 27.02  & 0.86\\
            \hline
        \end{tabular}
    \end{center}
\caption{\label{tab:1}Imaging quality on CIFAR-10 test set.}
\end{table}

Finally, since the CIFAR-10 dataset is usually used for the classification task, the classification experiments have also been done. We use the reconstructed gray images from scattering to train and test the ResNet20 classification model. The reconstructed gray images from the proposed LISMU-FCN and LISMU-OCN models can be classified with 72.40\% and 79.41\% accuracy, respectively. The accuracy of the gray images from original CIFAR-10 dataset (the colorful image are converted to gray images) is 89.31\%, which reflects the good imaging quality of the proposed DNNs model.}

As shown in Table \ref{tab:1}, after training, the average gray difference, $PSNR$, $SSIM$ and the classification accuracy, all indicate that DNNs have achieved good imaging results. Based on these results, both LISMU-FCN and LISMU-OCN perform well on imaging through scattering media. Furthermore, we can also draw a conclusion, whereby the imaging quality of LISMU-OCN is slightly better than that of LISMU-FCN.

\subsection{Analysis of Balance Loss Function}

\begin{figure}[htbp]
    \begin{minipage}[t]{0.45\textwidth}
        \centering
        \includegraphics[width=0.98\textwidth]{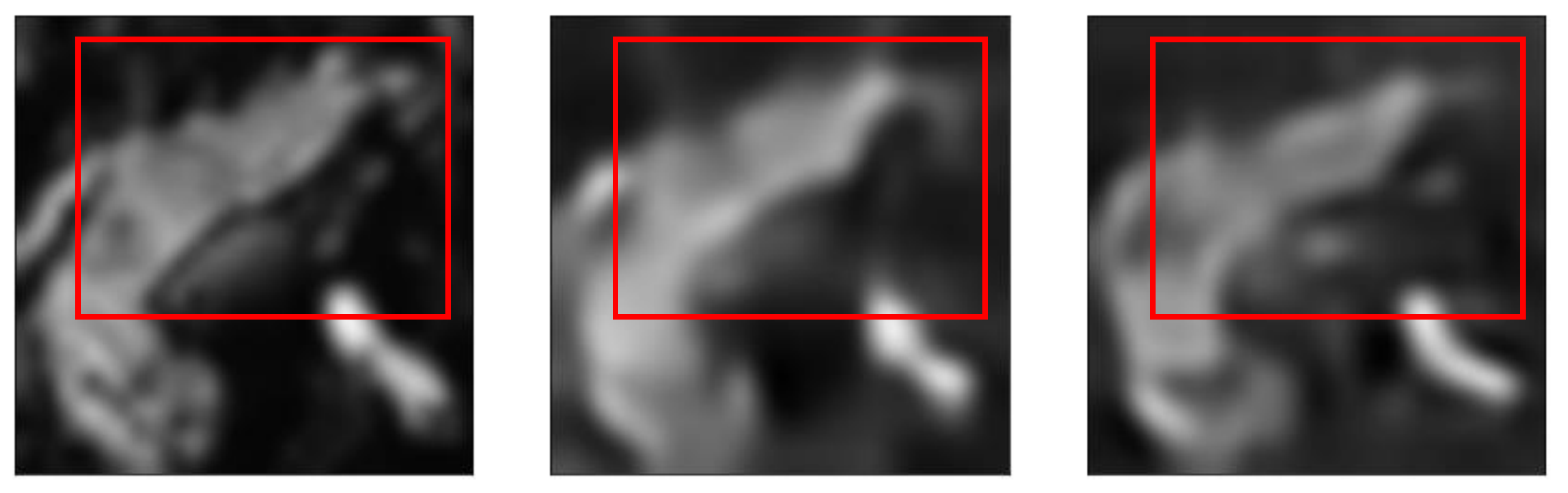}
        \caption{Comparison of output images from LISMU-FCN with and without balance loss. From left to right: ground truth (CIFAR-10\cite{CIFAR10_dataset}), LISMU-FCN output without balance loss, and LISMU-FCN output with balance loss.}
        \label{fig:8}        
    \end{minipage}
    \hspace{10pt}
    \begin{minipage}[t]{0.45\textwidth}
        \centering
        \includegraphics[width=0.98\textwidth]{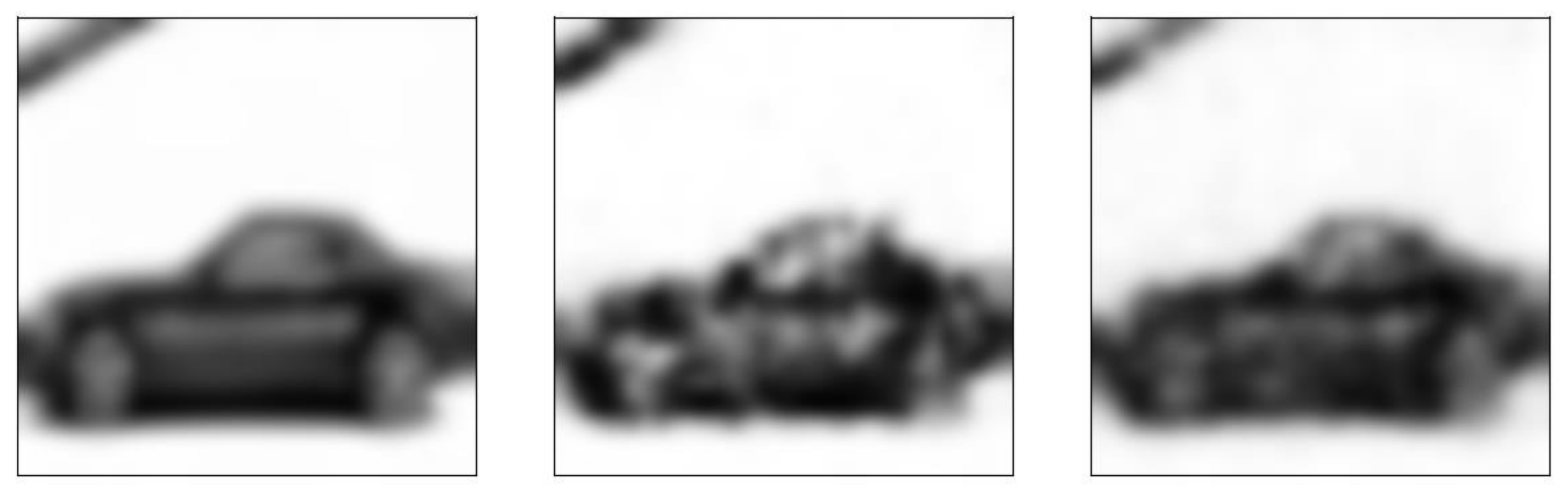}
        \caption{ Comparison of output images from LISMU-OCN with and without balance loss. From left to right: ground truth (CIFAR-10\cite{CIFAR10_dataset}), LISMU-OCN output without balance loss, and LISMU-OCN output with balance loss. }
        \label{fig:9}
    \end{minipage}    
    \vspace{-15pt}
\end{figure}

\bigskip
Due to the convolution in the final layer of LISMU-FCN, the sharpness and details of the output images are affected. Similarly, output images from LISMU-OCN contain more noise due to the final fully connected layer. To balance the image sharpness and smoothness, the balance loss term is used during training. Positive $\lambda$ in balance loss is used in LISMU-OCN for less noise, and negative $\lambda$ is used in LISMU-FCN for sharpness. Figure \ref{fig:8} and \ref{fig:9} illustrate that the output images of LISMU-FCN trained with the balance loss term are able to retain more texture details while output images of LISMU-OCN trained with balance loss term are less noisy.

\subsection{Transfer Learning on Other Datasets}\label{sec:exp-transfer}
\subsubsection{Distributions of Different Datasets}

\begin{figure}[htbp]
    \begin{minipage}[t]{0.45\textwidth}
        \centering
        \includegraphics[height=0.6\textwidth]{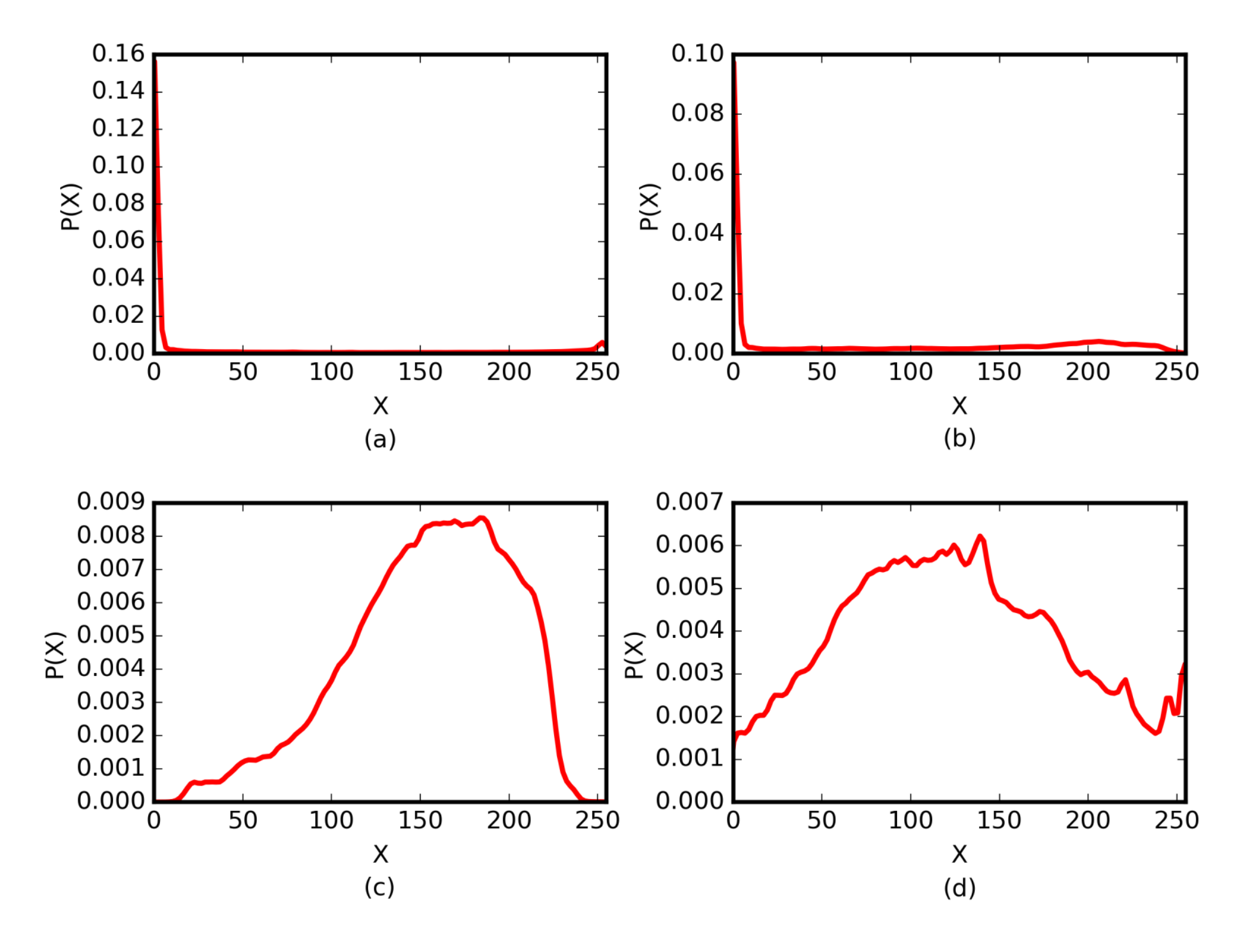}
        \caption{Gray value distribution of different datasets: (a) MNIST \cite{lecun1998gradient} dataset; (b) Fashion-MNIST \cite{xiao2017fashion} dataset; (c) Face \cite{1994Parameterisation} dataset (dataset obtained from AT\&T Laboratories Cambridge); (d) CIFAR-10 \cite{CIFAR10_dataset} dataset.}
        \label{fig:10}        
    \end{minipage}
    \hspace{10pt}
    \begin{minipage}[t]{0.45\textwidth}
        \centering
        \includegraphics[height=0.6\textwidth]{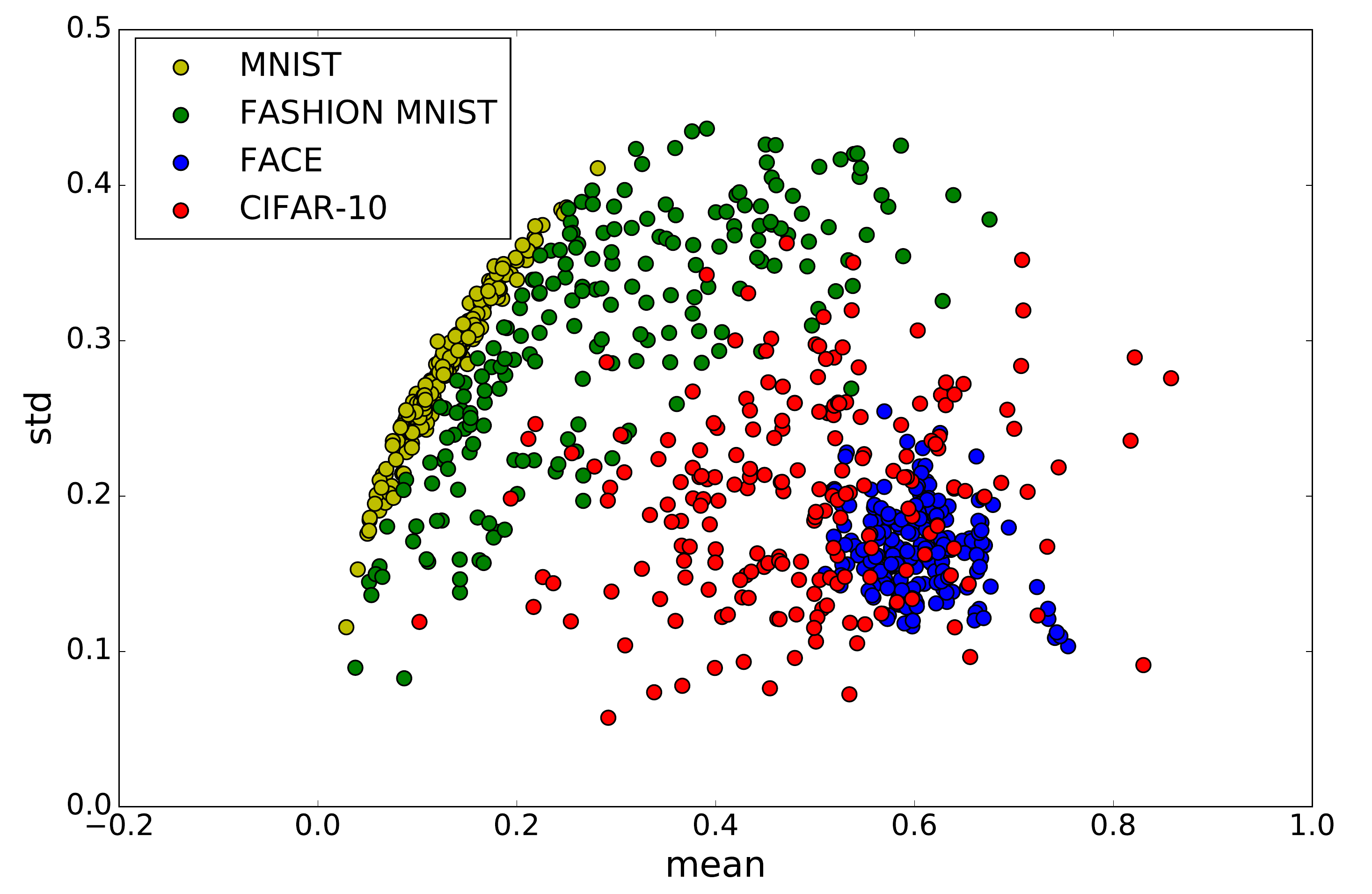}
        \caption{Mean-variance distributions of different datasets. }
        \label{fig:11}
    \end{minipage}    
    
\end{figure} 

While we expect to choose a relatively more complex dataset for model training, it would be good to quantify the distribution of pixel values in the training and other datasets so as investigate their differences in distributions (if any) and to justify the choice of the dataset for model training. We first analyze the distributions of four typical datasets, namely MNIST, Fashion-MNIST, Face, and CIFAR-10 dataset. Choosing 200 samples from each dataset, we compute the probability of occurrence of different gray values in a pixel and plot the distribution as shown in Figure \ref{fig:10}. We also compute the mean and variance of gray values of pixels in each image and plot them in a scatterplot as shown in Figure \ref{fig:11}.

{From Figure \ref{fig:10}, we can see that the gray value distribution on the CIFAR-10 dataset is more evenly distributed, which reflects this dataset is more general and relatively complex in one way.} And it really helps the DNNs model learned from the complex CIFAR-10 dataset more close to the real reconstructed model and makes it a suitable dataset for model training, compared to the other datasets. Figure \ref{fig:11} also confirms this point. The values of pixels in CIFAR-10 images are more widely distributed, while the Face dataset is also quite similar. MNIST and Fashion-MNIST dataset in comparison are more narrowly distributed with a mean close to zero.

\subsubsection{Vanilla Transfer Learning}

\begin{figure}[!htbp]
    \centering
    \includegraphics[width=0.84\textwidth]{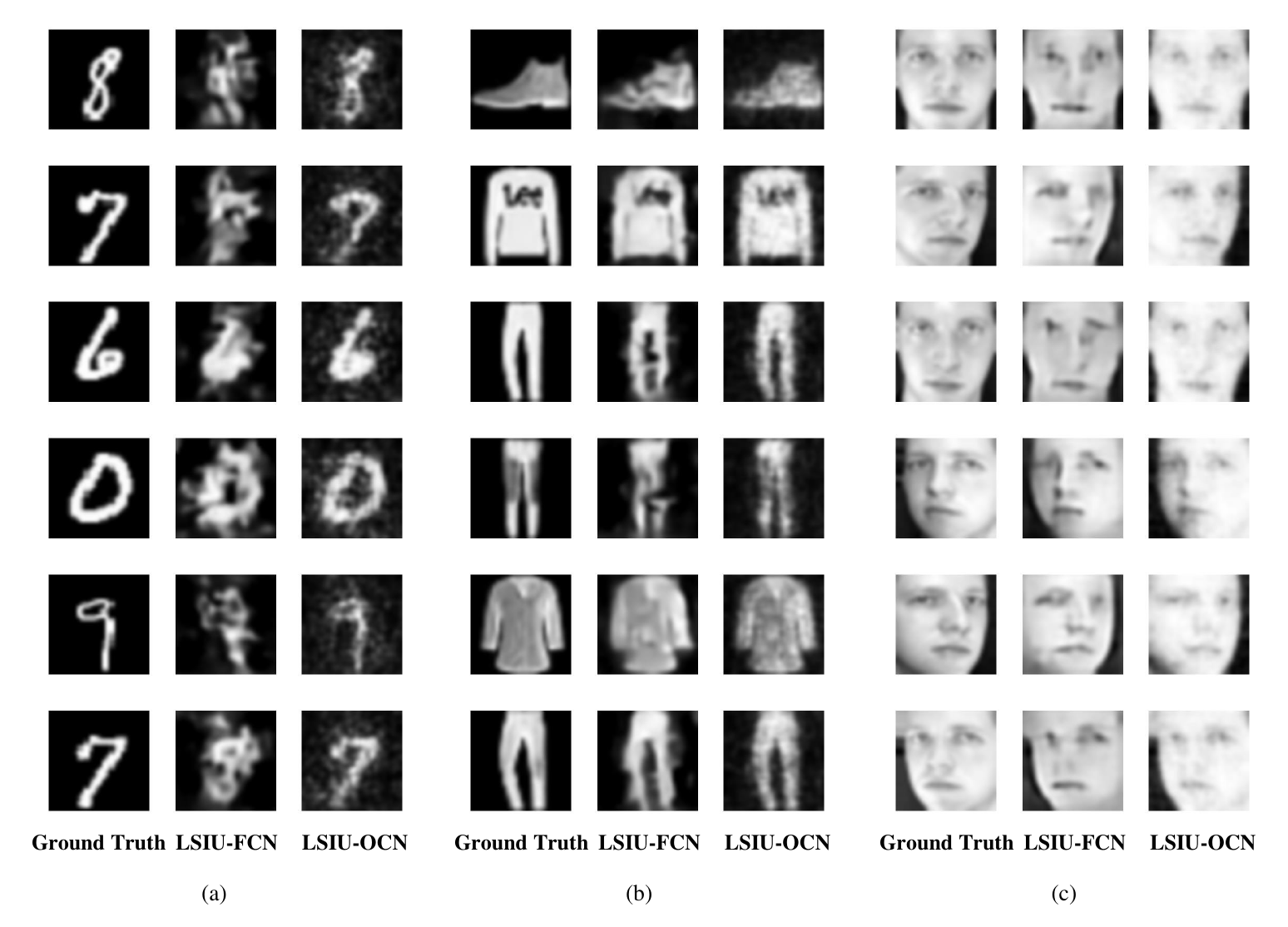}
    \caption{\label{fig:12} Vanilla transfer learning across different datasets with ground truth images, LISMU-FCN output images, LISMU-OCN output images on: (a) MNIST \cite{lecun1998gradient} dataset; (b) Fashion-MNIST \cite{xiao2017fashion} dataset; (c) Face \cite{1994Parameterisation} dataset (dataset obtained from AT\&T Laboratories Cambridge).}
\end{figure}

As CIFAR-10 is a more general dataset, we think that a model trained on CIFAR-10 would serve better as a baseline model for transfer learning. We test the trained model on MNIST, Fashion-MNIST, and Face datasets. Figure \ref{fig:12} shows that LISMU-OCN has slightly better transfer learning capability on the MNIST dataset and both models achieve relatively good imaging for the other datasets except MNIST. This is due to the huge difference in pixel value distributions between CIFAR-10 and MNIST.

 The average gray difference and PSNR in Table \ref{tab:2}, however, seem to suggest that the imaging quality for the Face dataset, whose distribution is much closer to CIFAR-10, is just slightly better than MNIST and worse than Fashion-MNIST. This is due to batch normalization\cite{Ioffe2015} being used in our CNN models. By addressing internal covariate shift, batch normalization allows simpler parameter initialization and faster training convergence and is used extensively in convolutional neural networks. In the process of testing the models on different datasets, the mean and variance of CIFAR-10 speckle intensity images are factored into the batch normalization layer and the mean and variance of MNIST and Fashion-MNIST speckle intensity images are much closer to those of CIFAR-10 speckle intensity images as compared to Face speckle intensity images. This causes the discrepancy between actual output images and the measures used. We can also see this point from the SSIM indicator.

\begin{table}[htbp]
\begin{center}
        \begin{tabular}{|c|c|c|c|c|c|c|} 
            \hline
            & \multicolumn{3}{|c|}{LISMU-FCN} & \multicolumn{3}{|c|}{LISMU-OCN}   \\
            \hline
            Dataset & $|\bigtriangleup x|$  & $PSNR(dB)$ & $SSIM$ & $|\bigtriangleup x|$&  $PSNR(dB)$ & $SSIM$    \\
            \hline
            MNIST & 39.58 & 12.15 & 0.29 &  37.88 & 14.85 & 0.40 \\
            \hline
            Fashion-MNIST & 16.20 & 20.24 & 0.60 & 22.07  & 19.84 & 0.58 \\
            \hline
            Face & 49.18 & 13.72 & 0.81 & 40.56  & 15.47 & 0.83 \\
            \hline
        \end{tabular}
    \end{center}
\caption{\label{tab:2}Imaging quality evaluation under vanilla transfer learning across different datasets.}
\end{table}

\subsubsection{Improved Transfer Learning through fine-tuning}
\begin{figure}[htbp]
    \centering
    \includegraphics[width=0.84\textwidth]{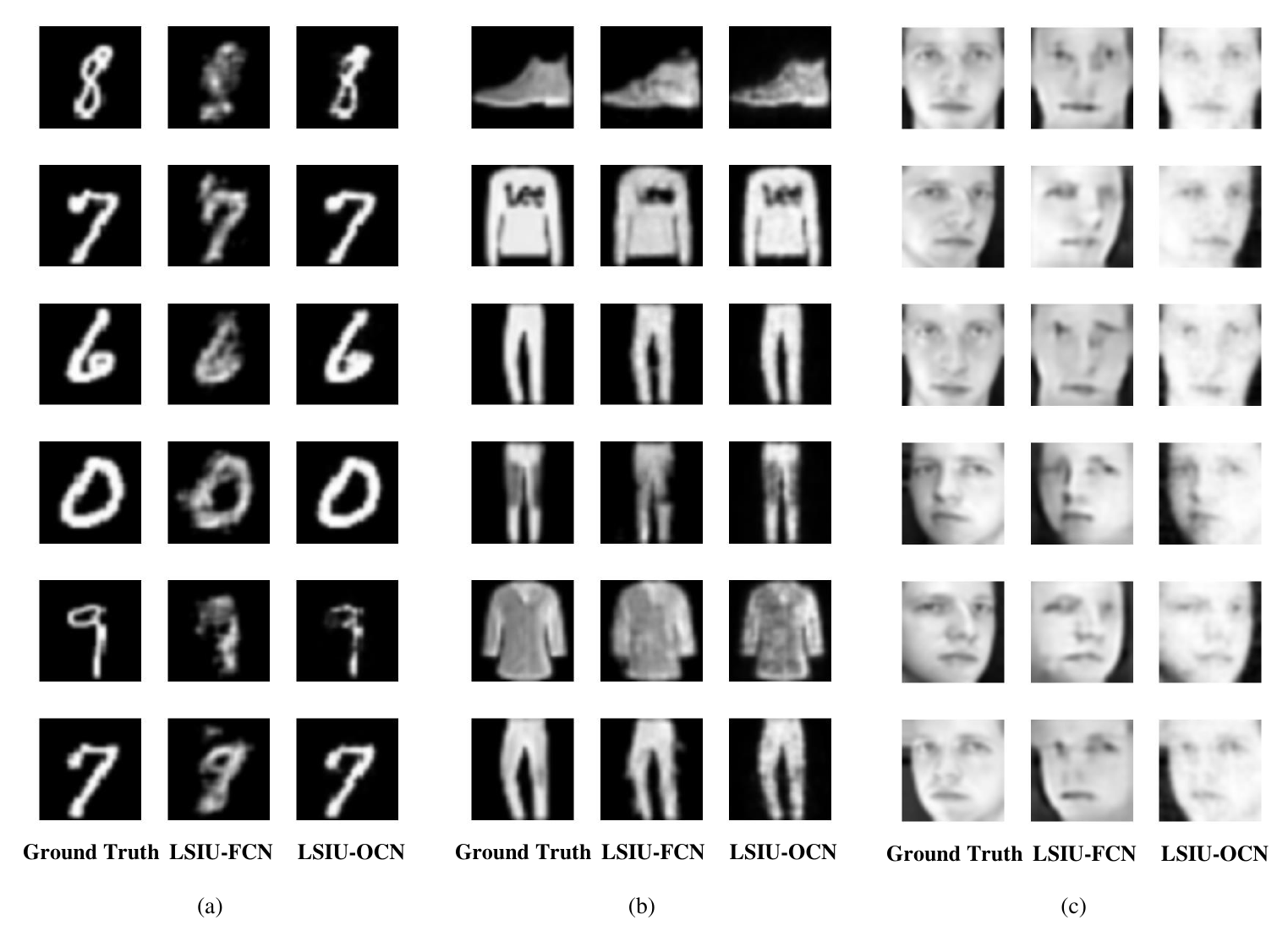}
    \caption{\label{fig:13}Improved transfer learning through model fine-tuning across different datasets with ground truth images, LISMU-FCN output images, LISMU-OCN output images on: (a) MNIST \cite{lecun1998gradient} dataset; (b) Fashion-MNIST \cite{xiao2017fashion} dataset; (c) Face \cite{1994Parameterisation} dataset (dataset obtained from AT\&T Laboratories Cambridge).}
\end{figure}

In order to further improve the transfer learning capability of the models, fine-tuning is performed. Other than the 50,000 images from CIFAR-10 that are used to train LISMU-FCN and LISMU-OCN, 300 images from each dataset (MNIST, Fashion-MNIST and Face) are used to fine-tune the models. During fine-tuning, only the last two layers of LISMU-FCN and LISMU-OCN are trained. Experiment results are presented in Figure \ref{fig:13}. It can be seen that, after fine-tuning, both LISMU-FCN and LISMU-OCN achieve much better imaging on all three datasets.

\begin{table}[!htbp]  
    \begin{center}
        \begin{tabular}{|c|c|c|c|c|c|c|} 
            \hline
            & \multicolumn{3}{|c|}{LISMU-FCN} & \multicolumn{3}{|c|}{LISMU-OCN}   \\
            \hline
            Dataset & $|\bigtriangleup x|$  & $PSNR(dB)$ & $SSIM$ & $|\bigtriangleup x|$&  $PSNR(dB)$ & $SSIM$   \\
            \hline
            MNIST & 20.02 & 14.95 & 0.58 &  6.86 & 22.08 & 0.61 \\
            \hline
            Fashion-MNIST & 11.12 & 22.19 & 0.68 & 9.36  & 25.13 & 0.70 \\
            \hline
            Face & 7.35 & 28.45 & 0.92 & 7.06  & 29.13 & 0.93 \\
            \hline
        \end{tabular}
    \end{center}
    \caption{\label{tab:3}Imaging quality evaluation under improved transfer learning through fine-tuning across different datasets.}
\end{table}

Table \ref{tab:3} demonstrates the imaging quality under improved transfer learning with fine-tuning. Compared to Table \ref{tab:2} under vanilla transfer learning, the imaging quality is greatly improved for both LISMU-FCN and LISMU-OCN, especially on MNIST. Although the imaging quality of LISMU-FCN is slightly worse than LISMU-OCN, it is still acceptable considering it has much fewer parameters. With less noise, the imaging quality of LISMU-FCN on the Face dataset (of similar distribution as CIFAR-10) is almost the same as LISMU-OCN.

{\subsubsection{Training Dataset Selection}

\begin{figure*}[htbp]
    \centering
    \includegraphics[width=1.0\textwidth]{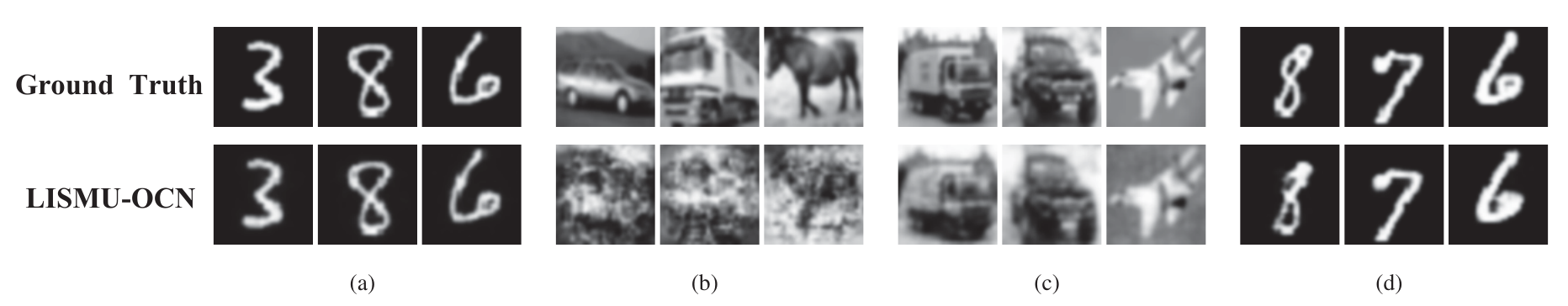}
    \caption{{Ground truth images and LISMU-OCN output images over different training and fine-tuning datasets: (a) training dataset: MNIST \cite{lecun1998gradient}, fine-tuning dataset: MNIST \cite{lecun1998gradient}; (b) training dataset: MNIST \cite{lecun1998gradient}, fine-tuning dataset: CIFAR-10 \cite{CIFAR10_dataset}; (c) training dataset: CIFAR-10\cite{CIFAR10_dataset}, fine-tuning dataset: CIFAR-10\cite{CIFAR10_dataset}; (d) training dataset: CIFAR-10\cite{CIFAR10_dataset}, fine-tuning dataset: MNIST \cite{lecun1998gradient}.}}
    \label{fig:14}
\end{figure*}

\begin{table}[htbp]    
    
    \begin{center}
        \begin{tabular}{|c|c|c|c|c|c|c|c|}
            \hline
            \multicolumn{2}{|c|}{\multirow{3}{*}{}} &
            \multicolumn{6}{|c|}{Fine-tuning dataset} \\
            \cline{3-8}
             \multicolumn{2}{|c|}{} & \multicolumn{3}{|c|}{MNIST} & \multicolumn{3}{|c|}{CIFAR-10}   \\
            \cline{3-8}
             \multicolumn{2}{|c|}{} & $|\bigtriangleup x|$  & $PSNR(dB)$ & $SSIM$ &   $|\bigtriangleup x|$&  $PSNR(dB)$ & $SSIM$    \\
            \hline
            \multirow{2}{*}{Training Dataset}
             & MNIST & 7.79 & 24.57 & 0.62 &  \bf{29.42} & \bf{16.83} &  \bf{0.39} \\
            \cline{2-8}
            & CIFAR-10 & \bf{6.86} & \bf{22.08} & \bf{0.61} &  9.34 & 27.02 & 0.86 \\
            \hline
        \end{tabular}
    \end{center}
    \caption{\label{tab:DS} {Imaging quality evaluation over different training and fine-tuning datasets.}}
\end{table}

To verify the idea that the relatively complex images with more texture details will make the trained model closer to the real imaging model, and then enhance the transfer learning ability and make it easy to image on the simple dataset, we train the same DNN model (LISMU-OCN) on the CIFAR-10 dataset and MNIST dataset and then fine-tune the model on the MNIST and CIFAR-10 dataset, respectively. The experiment results are shown in Table \ref{tab:DS} and Figure \ref{fig:14}. 

From Table \ref{tab:DS}, Figure \ref{fig:14}(a) and \ref{fig:14}(c), we can see that the models trained on MNIST and CIFAR-10 can both work well facing the same dataset. It can also be seen from Figure \ref{fig:14}(a), \ref{fig:14}(d) and evaluation indicators that the model trained on the CIFAR-10 dataset and then fine-tuned on the only 300 samples from the MNIST dataset can even achieve almost the same imaging quality as that of the model trained on the whole MNIST dataset with 50,000 samples. However, the model trained on the whole MNIST dataset and then fine-tuned on the 300 samples from the CIFAR-10 dataset can't achieve a satisfactory imaging quality.

These experimental results evidence our idea well. The training dataset selection is of great importance for transfer learning in the field of lensless imaging through scattering media. The relatively complex samples will better help the model to learn and then enhance the transfer learning ability.
}

\section{Conclusion}
In this research, we employ CNNs for the generation of images in the problem of lensless imaging through scattering media. It is a high dimensional inverse regression problem, with an analytical solution generally difficult to obtain. Also, the knowledge learned on one dataset is generally not transferable to other datasets, hence incurring huge costs in data acquisition and training from scratch for new datasets. To this end, transfer learning is proposed. By training the model on a relatively more complex dataset, we may then use far fewer samples from the new datasets to fine-tune the model, which translates to significant savings. {We have also analyzed the training dataset selection in transfer learning and its importance for imaging quality.} Specifically, two CNN architectures, namely, LISMU-FCN and LISMU-OCN are proposed, and a balance loss function is formulated to balance between smoothness and sharpness. LISMU-FCN can achieve real-time imaging across similar datasets with much fewer parameters; while LISMU-OCN has better transfer learning capability, allowing it to be used for datasets that differ substantially from the training dataset. {Moreover, a set of simulation algorithms is established which can speed up the research of lensless imaging through scattering media greatly.} This paper demonstrates a new DNNs solution for lensless imaging through scattering media with transfer learning capability.



\bibliographystyle{elsarticle-num}

\bibliography{library}

\end{document}